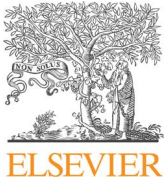
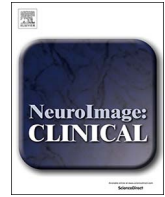

# The impact of epilepsy surgery on the structural connectome and its relation to outcome

Peter N. Taylor[a,b,c,*], Nishant Sinha[a,b], Yujiang Wang[a,b,c], Sjoerd B. Vos[d,e], Jane de Tisi[c], Anna Miserocchi[c], Andrew W. McEvoy[c], Gavin P. Winston[c,e,1], John S. Duncan[c,e,1]

[a] Interdisciplinary Computing and Complex BioSystems Group, School of Computing Science, Newcastle University, UK
[b] Institute of Neuroscience, Faculty of Medical Science, Newcastle University, UK
[c] NIHR University College London Hospitals Biomedical Research Centre, UCL Institute of Neurology, Queen Square, London WC1N 3BG, UK
[d] Translational Imaging Group, Centre for Medical Image Computing, University College London, UK
[e] Chalfont Centre for Epilepsy, Chalfont St Peter SL9 0LR, UK

ARTICLE INFO

ABSTRACT

*Background:* Temporal lobe surgical resection brings seizure remission in up to 80% of patients, with long-term complete seizure freedom in 41%. However, it is unclear how surgery impacts on the structural white matter network, and how the network changes relate to seizure outcome.
*Methods:* We used white matter fibre tractography on preoperative diffusion MRI to generate a structural white matter network, and postoperative T1-weighted MRI to retrospectively infer the impact of surgical resection on this network. We then applied graph theory and machine learning to investigate the properties of change between the preoperative and predicted postoperative networks.
*Results:* Temporal lobe surgery had a modest impact on global network efficiency, despite the disruption caused. This was due to alternative shortest paths in the network leading to widespread increases in betweenness centrality post-surgery. Measurements of network change could retrospectively predict seizure outcomes with 79% accuracy and 65% specificity, which is twice as high as the empirical distribution. Fifteen connections which changed due to surgery were identified as useful for prediction of outcome, eight of which connected to the ipsilateral temporal pole.
*Conclusion:* Our results suggest that the use of network change metrics may have clinical value for predicting seizure outcome. This approach could be used to prospectively predict outcomes given a suggested resection mask using preoperative data only.

## 1. Introduction

Epilepsy is a serious neurological disorder characterised by recurrent unprovoked seizures affecting 1% of the population. Neurosurgical resection can bring remission in up to 80% of those with refractory focal epilepsy, with 41% remaining entirely seizure free for years (De Tisi et al., 2011). The most common type of epilepsy surgery is anterior temporal lobe resection, in which the amygdala, anterior hippocampus, and anterior temporal neocortex are removed. The commonest neurological sequelae of temporal lobe surgery are memory impairment, visual field deficits and word-finding difficulties (Jutila et al., 2002; Gooneratne et al., 2017).

Recent studies have investigated surgical outcome by considering the brain as a network of connected regions. Such networks can then be subjected to quantitative analysis techniques, which measure local and global properties in networks (see Bernhardt et al. (2015) for review). Network measures that have been found to be altered in temporal lobe epilepsy (TLE) include the clustering coefficient of a region, which captures the connectedness of neighbours of a region (Bernhardt et al., 2011). Furthermore, the strength of a connection (e.g. the number of streamlines connecting two areas), or the strength of a region's connectivity (e.g. the number of streamlines connecting a region to all other regions) may also be altered in TLE (Besson et al., 2014a, 2014b; Taylor et al., 2015). Another measure of a network is its efficiency, which is a measure of network integration - i.e. how easy it is to travel between one region to another via direct and indirect paths, and has been shown to be altered in patients with TLE (Liu et al., 2014). Finally, regression analysis and machine learning approaches have also been applied to brain networks of TLE to relate them to surgical outcome (Bonilha et al., 2013; Munsell et al., 2015; Bonilha et al., 2015; Ji et al., 2015).

A challenge in comparing networks across subjects is the choice of an appropriate baseline or benchmark. There are two common approaches to this. One is to threshold the connectivity so that all subjects have the same number of connections. A range of thresholds are then checked, and the most significant results reported across thresholds (Zhang et al., 2011). This has the drawback of removing 'weak' but potentially important connections. A second approach is to compare the network to a random network with the same number of regions and connections. Typically this is done by either rewiring the existing network (Maslov and Sneppen, 2002), or by generating a new network according to predefined rules (Betzel et al., 2016; Bauer and Kaiser, 2017). Many different types of baseline networks can be used and this will therefore influence results.

Recently Kuceyeski et al. (2013) introduced the network modification (NeMo) tool in the context of stroke (Kuceyeski et al., 2015a, 2015b) and multiple sclerosis (Kuceyeski et al., 2015a, 2015b). The NeMo tool is a method to enable a direct comparison between networks that undergo change. For example, in their study of stroke, the authors drew masks over stroke affected areas and overlaid this mask with data from healthy subjects. Normal connectivity from the healthy subjects, and altered connectivity (i.e. tracts which pass through the stroke mask) were calculated. This approach allowed the authors to calculate a change in connectivity metric (ChaCo), which was shown to correlate with outcomes. Since the authors use the pre-stroke network as a baseline to investigate the implied post-stroke differences, the analysis is possible without the need to generate random networks or threshold the connectivities. This obviates the need for arbitrarily chosen surrogate networks by effectively using the patient's own network as the surrogate instead – a distinct advantage of the technique. A drawback of that study is that the tractography was derived from a cohort of healthy controls, rather than the stroke patients. Nonetheless, this framework is ideally suited to investigate *changes* in networks, given a well-defined alteration such as a stroke or surgery.

In this study we used a ChaCo-like approach in the context of epilepsy surgery and addressed the following questions: What is the impact of surgery on the patient's network? How does this impact graph theoretic properties such as region strength, network efficiency? Do these *changes* to patient networks correlate with surgical outcome?

Although the resection masks we use in this study are derived retrospectively from postoperative data, our methods could in future be applied preoperatively using a mask of the intended resection.

## 2. Materials and methods

### 2.1. Patients & MRI acquisition

We retrospectively studied 53 patients who underwent temporal lobe epilepsy surgery at the National Hospital for Neurology and Neurosurgery, London, United Kingdom. Full patient details can be found in Table S11, a summary is given in Table 1. Patient outcomes were defined at 12 months postoperatively, according to the ILAE classification of surgical outcomes (Wieser et al., 2001) and separated into two groups. Group 1 includes patients who were completely seizure free (ILAE 1), and group 2 incorporates all other possibilities (ILAE 2–6). No patient had any prior history of neurosurgery. We used a $\chi^2$ test to check for differences between outcome groups in gender, side of surgery, and evidence of hippocampal sclerosis. We applied Kruskal-Wallis test to check for differences in age between outcome groups.

All patients underwent preoperative anatomical T1-weighted MRI and preoperative diffusion MRI. Postoperative T1-weighted MRI was obtained within 12 months of surgery with the exception of one patient, who was rescanned later.

MRI studies were performed on a 3T GE Signa HDx scanner (General

Table 1
Patient demographics and relation to outcome group.

|  | ILAE 1 | ILAE 2–6 | Significance |
|---|---|---|---|
| N | 36 (68%) | 17 (32%) | |
| Male/female | 16/20 | 4/13 | $p = 0.3597$, $\chi^2 = 0.839$ |
| Left/right TLE | 22/14 | 8/9 | $p = 0.3353$, $\chi^2 = 0.923$ |
| Age (mean, S.D./median, I.Q.R.) | 37, 11.6/ 39.6, 19.25 | 41.5, 10.6/ 42.3, 10.8 | $p = 0.2374$ |
| Hippocampal sclerosis | 25 (69%) | 10 (59%) | $p = 0.4460$, $\chi^2 = 0.5808$ |

Electric, Waukesha, Milwaukee, WI). Standard imaging gradients with a maximum strength of $40\,\text{mT}\,m^{-1}$ and slew rate $150\,\text{T}\,m^{-1}\,s^{-1}$ were used. All data were acquired using a body coil for transmission, and 8-channel phased array coil for reception. Standard clinical sequences were performed including a coronal T1-weighted volumetric acquisition with 170 contiguous 1.1 mm-thick slices (matrix, 256 × 256; in-plane resolution, 0.9375 × 0.9375 mm).

Diffusion MRI data were acquired using a cardiac-triggered single-shot spin-echo planar imaging sequence (Wheeler-Kingshott et al., 2002) with echo time = 73 ms. Sets of 60 contiguous 2.4 mm-thick axial slices were obtained covering the whole brain, with diffusion sensitizing gradients applied in each of 52 noncollinear directions (b value of $1{,}200\,\text{mm}^2\,s^{-1}$ [$\delta = 21$ ms, $\Delta = 29$ ms, using full gradient strength of $40\,\text{mT}\,m^{-1}$]) along with 6 non-diffusion weighted scans. The gradient directions were calculated and ordered as described elsewhere (Cook et al., 2007). The field of view was 24 cm, and the acquisition matrix size was 96 × 96, zero filled to 128 × 128 during reconstruction, giving a reconstructed voxel size of 1.875 × 1.875 × 2.4 mm. The DTI acquisition time for a total of 3480 image slices was approximately 25 min (depending on subject heart rate).

### 2.2. Image processing

#### 2.2.1. T1 processing

Preoperative anatomical MRI was used to generate parcellated regions of interest (network nodes: ROIs). We used two different approaches to do this, generating two different parcellation schemes. First, we used the FreeSurfer recon-all pipeline (https://surfer.nmr.mgh.harvard.edu/), which performs intensity normalization, skull stripping, subcortical volume generation, gray/white segmentation, and parcellation (Fischl, 2012). The default parcellation scheme from FreeSurfer (the Desikan-Killiany atlas (Fischl et al., 2002; Desikan et al., 2006)) contains 82 cortical ROIs and subcortical ROIs and is widely used in the literature (e.g. Munsell et al., 2015; Taylor et al., 2015). The method FreeSurfer uses to generate its ROIs uses anatomical priors based on a manually annotated dataset from healthy controls. However, this may be suboptimal in the case of disease and therefore, we use a second approach based on geodesic information flow (GIF) to generate ROIs which has the advantage of performing well even in the presence of neuropathology (Cardoso et al., 2015). Using the GIF approach, we generate 114 cortical and subcortical ROIs (Table 2). A drawback of using the GIF approach is comparison to previous studies is less straightforward since most previous work use alternative atlases. The results presented in the main manuscript use the GIF derived ROIs, while we include results using FreeSurfer derived ROIs in supplementary materials to aid comparison to previous studies.

#### 2.2.2. DWI processing

Preoperative diffusion MRI data were first corrected for signal drift (Vos et al., 2016), then eddy current and movement artefacts were corrected using the FSL eddy_correct tool (Andersson and Sotiropoulos, 2016). The b vectors were then rotated appropriately using the 'fdt-rotate-bvecs' tool as part of FSL (Jenkinson et al., 2012; Leemans and Jones, 2009). The diffusion data were reconstructed using generalized q-sampling imaging (Yeh et al., 2010) with a diffusion sampling length ratio of 1.2. A deterministic fibre tracking algorithm (Yeh et al., 2013) was then used, allowing for crossing fibres within voxels, with seeds placed at the whole brain. Probabilistic approaches to fibre tracking (e.g. Tournier et al., 2012) have been shown to perform less well with respect to accuracy and number of false positive connections inferred.[2] The choice of tractography algorithm is therefore related to whether sensitivity or specificity is more important. It was recently shown (Zalesky et al., 2016) that the introduction of false positive connections is substantially more detrimental to the calculation of graph theoretic measures such as efficiency and clustering than the introduction of false negatives. We therefore use deterministic tractography since this has been shown to have fewer false positive connections[1]. Default tractography parameters from the 14 February 2017 build of DSI studio software were used as follows. The angular threshold used was 60 degrees and the step size was set to 0.9375 mm. The anisotropy threshold was determined automatically by DSI Studio. Tracks with length < 10 mm and > 300 mm were discarded. A total of 1,000,000 tracts were calculated per subject and saved in diffusion space.

Table 2
Full names of the abbreviated regions of interest (ROIs).

| Full name | Abbreviation | Full name | Abbreviation |
| --- | --- | --- | --- |
| Anterior cingulate gyrus | ACgG | Occipital pole | OCP |
| Anterior insula | AIns | Occipital fusiform gyrus | OFuG |
| Anterior orbital gyrus | AOrG | Opercular part of the inferior frontal gyrus | OpIFG |
| Angular gyrus | AnG | Orbital part of the inferior frontal gyrus | OrIFG |
| Calcarine cortex | Calc | Posterior cingulate gyrus | PCgG |
| Central operculum | CO | Precuneus | PCu |
| Cuneus | Cun | Parahippocampal gyrus | PHG |
| Entorhinal area | Ent | Posterior insula | PIns |
| Frontal operculum | FO | Parietal operculum | PO |
| Frontal pole | FRP | Postcentral gyrus | PoG |
| Fusiform gyrus | FuG | Posterior orbital gyrus | POrG |
| Gyrus rectus | GRe | Planum polare | PP |
| Inferior occipital gyrus | IOG | Precentral gyrus | PrG |
| Inferior temporal gyrus | ITG | Planum temporale | PT |
| Lingual gyrus | LiG | Subcallosal area | SCA |
| Lateral orbital gyrus | LOrG | Superior frontal gyrus | SFG |
| Middle cingulate gyrus | MCgG | Supplementary motor cortex | SMC |
| Medial frontal cortex | MFC | Supramarginal gyrus | SMG |
| Middle frontal gyrus | MFG | Superior occipital gyrus | SOG |
| Middle occipital gyrus | MOG | Superior parietal lobule | SPL |
| Medial orbital gyrus | MOrG | Superior temporal gyrus | STG |
| Postcentral gyrus medial segment | MPoG | Temporal pole | TMP |
| Precentral gyrus medial segment | MPrG | Triangular part of the inferior frontal gyrus | TrIFG |
| Superior frontal gyrus medial segment | MSFG | Transverse temporal gyrus | TTG |
| Middle temporal gyrus | MTG | | |

### 2.2.3. Resection mask

To draw resection masks we linearly registered the postoperative T1 MRI to the preoperative MRI using the FSL FLIRT tool (Jenkinson and Smith, 2001; Jenkinson et al., 2002). Resection masks were then manually drawn using the FslView software by overlaying the post-operative MRI with the preoperative MRI starting at the most anterior coronal slice, then proceeding posteriorly every three slices. Attention was given to ensure masks did not extend beyond the Sylvian fissure into inferior frontal lobe since this is unrealistic for anterior temporal resection and there was no evidence for this on post-operative MRI. Once complete, coronal slices were then joined by drawing in every sagittal slice. Masks were saved in the preoperative T1 space. Linear registration is justified in this instance because nonlinear deformation of the tissue around the resection site may lead to misleading boundaries due to local deformations induced as part of the processing. All registrations were visually inspected in detail during the drawing of the resection mask.

### 2.3. Network generation

To align the tracts with the ROIs we linearly registered the preoperative T1 image to the first non-diffusion-weighted image and saved the transformation matrix using FSL FLIRT. We then multiplied every coordinate in every tract by the inverse of this transformation matrix to get the tracts in T1 space. The quality of the registration between tracts, ROIs, and the resection mask was confirmed through visual inspection for all subjects. Since networks are constructed in native space, this removes any mismatching of track types due to potential nonlinear registration issues which is advantageous compared to previous studies of network change (Kuceyeski et al., 2016). To generate preoperative connectivity matrices, we looped through all tracts and deemed two regions as connected if the two endpoints of the tract terminate in those regions. This generated a weighted connectivity (adjacency) matrix in which each entry in the matrix represents the number of streamlines connecting two regions. To generate predicted postoperative connectivity matrices we performed the same process as above with one exception. Any tract that had any point within the resection mask was excluded from building the matrix. The inferred postoperative network therefore always had fewer streamlines than the preoperative network and makes the assumption that remaining portions of removed tracts subserve no functionality. This reduction in streamlines alone did not explain outcome (Fig. S1). For analysis we applied a $\log_{10}$ transformation to connection weights. Following this, right hemisphere regions (rows and columns in the matrix) were flipped in patients with a right sided resection as we investigated ipsi- and contralateral differences. All subjects had 72 regions per hemisphere.

A summary of the processing pipeline to generate the GIF connectivity matrices is shown in Fig. 1.

### 2.4. Visualisation

To visualise the resection mask we first linearly registered the preoperative T1 to the MNI brain template using FSL FLIRT to generate a transformation matrix. Next, we nonlinearly registered the preoperative T1-weighted image to the MNI template using FSL FNIRT initialised with the aforementioned transformation to generate a nonlinear warp. Finally, we applied this warp to both the preoperative T1-weighted image, and mask. All images were visually inspected to check for registration quality. We repeated this for all patients to generate a mask in MNI space. These masks were then loaded into MatLab, binarised (thresholded at 0.5), and summed across all subjects. This was then saved, and overlaid with the MNI brain using FSLView. Masks in MNI space were generated for visualisation purposes only and not used in the analysis.

Three-dimensional projections of brain regions were produced using the centre of mass for each region and visualised using BrainNet Viewer (Xia et al., 2013). We created the scatter plots using the UniVarScatter function in Matlab (https://github.com/GRousselet/matlab_visualisation).

---

[2] http://www.tractometer.org/downloads/downloads/ismrm_presentation/Challenge_ResultTractometer_updated_for_pdf_generation.pdf.

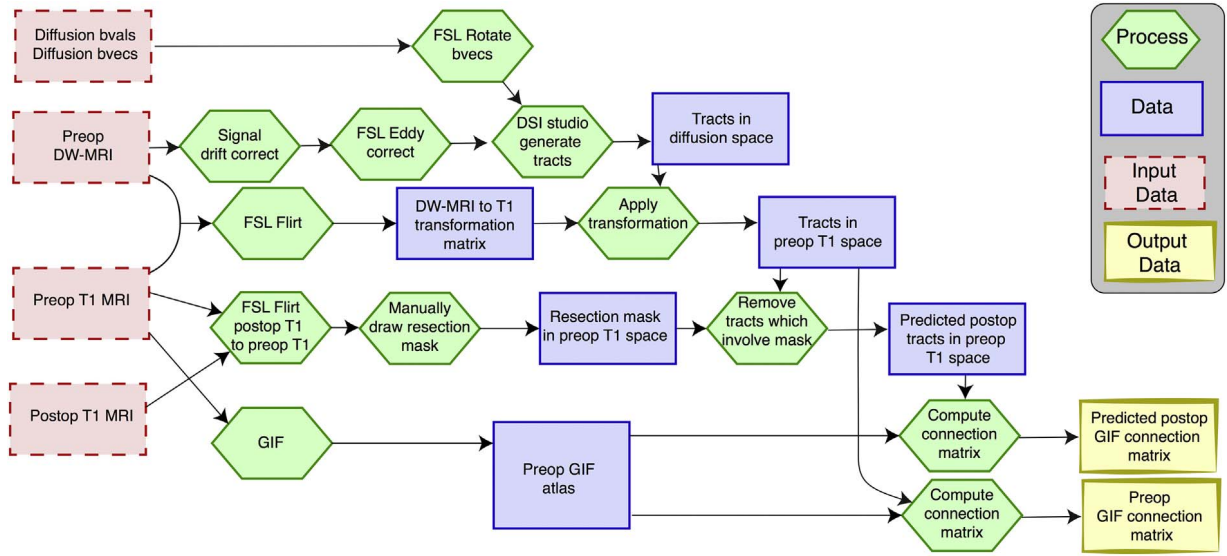

**Fig. 1. Processing pipeline summary for GIF matrix generation.**
The pipeline is applied to each subject.

### 2.5. Network analysis

We investigated properties of the networks using well established graph theory measures that assessed the properties of a region, the properties of a connection, and the properties of the network as a whole (i.e. global). We excluded self-connections from all measures while applying the weighted and undirected versions of the functions in the Brain Connectivity Toolbox (Rubinov and Sporns, 2010). We investigated measures of strength, regional communicability (sum of communicability matrix Estrada and Hatano, 2008), betweenness centrality, clustering, and global efficiency, since several of these have been suggested to be altered in previous studies (Bernhardt et al., 2011; Besson et al., 2014a, 2014b; Taylor et al., 2015; Liu et al., 2014).

Region volume, region strength, connection strength changes were computed by dividing the post-operative network property by the pre-operative network property. For example, if node strength reduces from 10 pre-operatively to 2 in the inferred post-operative network the change measurement will be 0.2. Other measures (betweenness, clustering) can increase or decrease after surgery and so the difference was computed by subtracting the pre-operative value from the post-operative value.

### 2.6. Elastic net regularisation and support vector machine classification framework

We implemented a machine learning framework to investigate the association of metrics indicating changes in the region strength, region volumes, and connection strength with seizure free (ILAE 1) and non-seizure free (ILAE 2–6) outcomes. We computed the aforementioned metrics and treated them as feature vectors. Specifically, the feature space included 380, 85, and 14 features computed from proportional reduction in the connection strength, region strength, and region volumes respectively for each subject. This resulted in a high-dimensional feature matrix, $\mathbf{A} \in \mathbb{R}^{n \times m}$, where $m = 479$ indicates the total number of features and $n = 53$ denotes the total number of subjects in our study.

Since the number of features, $m$, is much larger than the number of subjects, $n$, we implemented a two-step classification paradigm (Munsell et al., 2015; Casanova et al., 2011). In the first step, we performed feature selection by implementing logistic regression with elastic net regularisation (Zou and Hastie, 2005). This gives a sparse $m$-dimensional weight vector, $\mathbf{x} \in \mathbb{R}^m$, which minimises the following regularised logistic regression problem:

$$\min_{\mathbf{x}} \sum_{i=1}^{n} w_i \log(1 + e^{(-y_i(\mathbf{x}^T \mathbf{a_i} + c))}) + \frac{\rho}{2} \left\| \mathbf{x} \right\|_2^2 + \lambda \left\| \mathbf{x} \right\|_1 \quad (1)$$

where, $y = (y_1, y_2, \cdots, y_n)$ is the $n$-dimensional vector representing the surgical outcomes (+1 for seizure-free outcome and −1 for nonseizure-free outcome); $\mathbf{a_i}^T$ denotes the $i-$th row of feature vector $\mathbf{A} \in \mathbb{R}^{n \times m}$; $w_i$ is the weight for the $i-$th sample (all samples were equally weighted in our implementation); $c$ is the scalar intercept; $\lambda \| \mathbf{x} \|_1$ is the $l_1$ regularisation (sparsity) term; and $\frac{\rho}{2} \|\mathbf{x}\|_2^2$ is the $l_2$ regularisation (smoothness) term. We optimised the cost function in Eq. 1 by applying the LogisticR method in the sparse learning with efficient projections (SLEP) software package (Liu et al., 2009a, 2009b; Liu and Ye, 2009; Liu and Ye, 2009; Duchi et al., 2008).

The weight vector $\mathbf{x}$ computed from Eq. 1 is a sparse representation of the training data set, i.e., the zero weights indicate the features which are not associated with surgical outcome, whereas the non-zero weights indicate the features associated with the surgical outcome. We derived a binary mask by setting all the non-zero weights in $\mathbf{x}$ to 1. Finally, we applied this mask to the high-dimensional feature space while selecting only the features associated with surgical outcome, resulting in a low-dimensional feature space.

In the second step of our classification framework, we incorporated the aforementioned reduced feature representation in a support vector machine (SVM) classifier with a linear kernel. We designed the SVM classifier in MATLAB using the 'fitcsvm' class. Specifically, we divided our data samples into test and training sets, incorporated a leave-one-out cross-validation scheme, and computed the average accuracy, sensitivity, and specificity. To prevent the classifier from evaluating skewed classification performance due to the between-class imbalance—36 in the seizure-free class as opposed to 17 in the non-seizure-free class—we implemented a class-weighted SVM by setting all class prior probabilities as uniform similar to previous studies (e.g. Wang et al., 2012). This sets a higher misclassification penalty for the minority class as compared to the majority class, thereby preventing the classifier from learning simply from the underlying empirical data distribution (Osuna et al., 1997; Veropoulos et al., 1999; Wu and Rohini, 2004).

Since our objective was to determine a minimum set of features that would most accurately separate the seizure-free outcome and the nonseizure-free outcome classes, we incorporated a leave-one-out cross-validated grid search on the $l_1$ and $l_2$ regularisation parameters ($\lambda, \rho$) (Munsell et al., 2015; Casanova et al., 2011). We varied the $\lambda$ parameter sequentially from 0.01 to 0.96 in steps of 0.05 and the ρ parameter from 0.01 to 1.96 in steps of 0.05. At each grid point, we incorporated the machine learning framework detailed above, obtained a reduced feature space, and computed the performance metrics indicating the accuracy, sensitivity, and specificity.

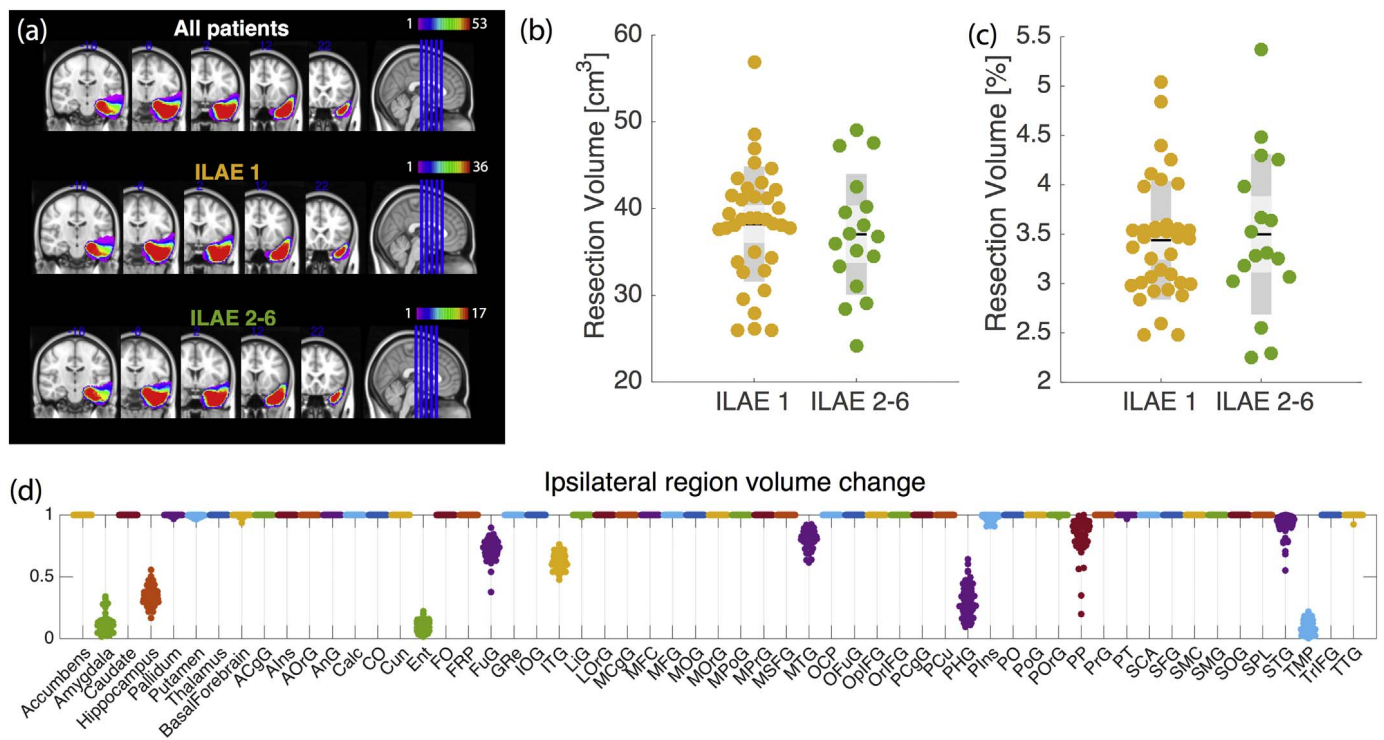

**Fig. 2.** Resection mask and region volume changes after surgery.
**(a)** Region mask in MNI space for visualisation only (all other analyses were performed in preoperative T1 space) shows similar spatial profile between outcome groups. **(b)** Absolute amount of tissue resected does not relate with outcome. **(c)** Percentage of whole brain removed, does not relate to outcome. **(d)** Proportion of tissue remaining after surgery varies between subjects and regions. Here, 1 represents that the region remained intact, 0 represents that region was removed entirely. Full region names can be found in Table 2 Each dot per region is a single subject (53 per region in total).

## 3. Results

Firstly, we investigate the inferred impact of surgery on network regions and secondly, on network connections. Finally, using machine learning we investigate how network change metrics relate to outcomes from surgery.

### 3.1. Impact of surgery on network regions

#### 3.1.1. Region volumes

All patients had similar surgical resections with the most variation between subjects on the boundary of the resection site (Fig. 2a). The absolute (Fig. 2b), and relative (Fig. 2c) volume of the resected tissue did vary between subjects, but did not explain outcome ($p = 0.56$ and $p = 0.62$ respectively – permutation test for difference between mean, 10,000 permutations used). Fig. 2d shows the proportion of tissue remaining after resection in different brain regions. Unsurprisingly the regions which were most reduced in volume were amygdala, hippocampus, entorhinal cortex, parahippocampal gyrus, and the temporal pole. This suggests partial disruption to multiple regions, rather than complete disruption to specific regions. Although there is variation between subjects in the proportion of each region removed, this did not predict outcome (Fig. S2). These results are reproduced for the FreeSurfer atlas in Fig. S3.

#### 3.1.2. Region strength

Each brain region is connected to other regions with streamlines.

The sum of all streamlines directly connected to a region is termed the region strength - i.e. how strongly a region is directly connected to all other regions. Fig. 3 shows the impact of removing all streamlines that pass into the resected tissue by plotting the resultant change in strength. As expected, those regions that are partially removed (i.e. those showing a reduction in Fig. 2d) also have reduced strengths. The effect of surgery on other brain regions can be demonstrated with this approach. For example, the ipsilateral basal forebrain, although it is extratemporal and not resected, has a substantial reduction in strength for many patients. This suggests that preoperatively the basal forebrain has many of its connections with the resected tissue, rather than other areas. Several other areas beyond the resected regions also undergo alterations in strength in many subjects including the ipsilateral lateral, medial and posterior, orbital gyri among others. These results are shown in detail in Table S12(a, b), and repeated for the FreeSurfer regions in Fig. S4. It is worth noting that region strength changes shown here may not be limited to only regions partly, or wholly resected, but may also change if a portion of a tract travels through the resection site, effectively disconnecting two non-resected areas.

Broadly speaking, the strengths of contralateral regions were not as substantially affected by the surgery which suggests most connections are intrahemispheric. None of the changes in node strength met significance for association with seizure-free outcome or not.

#### 3.1.3. Region network measures

The region volume change (Fig. 2) captures information solely about individual regions, while the region strength change (Fig. 3) captures information about a region and its directly connected neighbours. Other measures of network regions capture information regarding aspects of their role in the wider network.

The betweenness centrality of a region represents how many of the shortest paths between other regions traverse the region. Hub regions tend to occur in many of the shortest paths and have been implicated in a wide range of neurological disorders (Crossley et al., 2014). If a region is surgically removed, the shortest path between regions may be via other areas. The betweenness centrality of a region can therefore either increase or decrease following surgery.

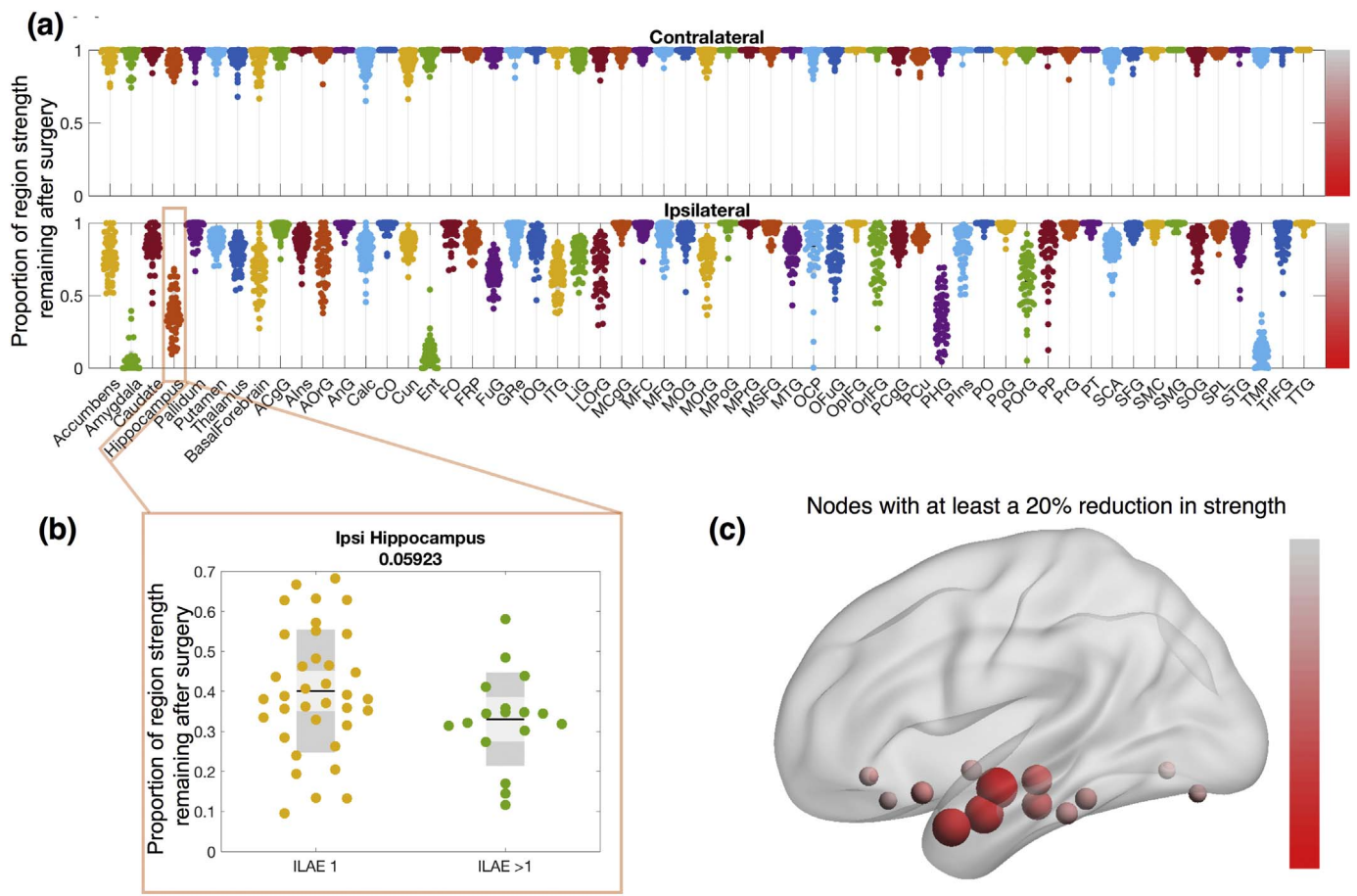

**Fig. 3.** Changes in region strength are predominantly ipsilateral.
**(a)** Proportion of strength remaining after surgery. A value of 1 (0) indicates all streamlines connecting to that region are kept (removed). **(b)** An example region shows ipsilateral hippocampal connection change in the ILAE > 1 group, and the ILAE 1 group. **(c)** Substantially affected regions visualised in 3D MNI space. Size and colour correspond to median reduction across all patients. Colour bars in panel (c) and (a) are consistent and correspond to the proportion of strength remaining between 0 and 1.

Fig. 4 shows the change in betweenness centrality after surgery. Regions involved in the resection have reductions in betweenness centrality suggesting reduced integration of those areas with the wider network. However, if those regions are no longer occurring in shortest paths, the following question remains: which ones are? Fig. 4a shows that many regions have substantial increase in betweenness centrality. The redirection of shortest paths in the network following surgery appears to be widely distributed among many other regions, rather than just the hubs - suggesting there are many alternative pathways (i.e. redundancy). Panel (b) in Fig. 4 shows the spatial distribution of the median change across all patients. This confirms that the increases (shown in blue) are widespread, even in contralateral regions.

Because of the widespread redistribution of the shortest paths, we also evaluated global efficiency, which is the inverse of the average shortest path length of the network. Fig. 4c demonstrates the reduction in global efficiency following surgery. Given the redirection of shortest paths shown in panel (a) of the same figure, it is now easy to see why the reduction on global efficiency is only < 10% in most cases. This change does not relate to seizure outcome ($p = 0.49$ – permutation test for difference between mean, 10,000 permutations used). We observed similar results when using the FreeSurfer atlas (Fig. S5).

We additionally investigated the change in regional clustering coefficient, a measure of the interconnectedness of neighbours of a region, following surgery. We found very subtle differences in this measure beyond the resection site following surgery (Fig. S6), which did not relate to outcome. Reductions of regional network communicability were most pronounced in ipsilateral amygdala, entorhinal, and temporal pole regions but also did not relate to outcome ($p > 0.05$, Kruskal-Wallis test, supplementary results S7).

### 3.2. Impact of surgery on network connections

#### 3.2.1. Connection strength

Fig. 5a shows the connections which are affected by surgery in all patients. These connections typically involve those areas which are resected. Panel (b) shows the median reduction in connectivity following surgery. Connections in panel (a) are a subset of those in panel (b). There are many connections which are only partly (as opposed to entirely) reduced in strength. Partial reduction in strength can be considered as removing some, but not all, of the connectivity between two areas. The greatest impact was seen in connections into the ipsilateral posterior temporal lobe and the inferior frontal lobe, with in some patients, connections to the ipsilateral parietal lobe and contralateral temporal lobe and medial frontal and parietal lobes as would be expected given Fig. 3.

#### 3.2.2. Connection network measures

Similar to the regional betweenness centrality, connection (edge) betweenness centrality captures information regarding how often a connection occurs in the shortest path between other regions – i.e. how frequently that connection is traversed in the shortest paths. This value can increase or decrease following surgery, as alternative paths is taken.

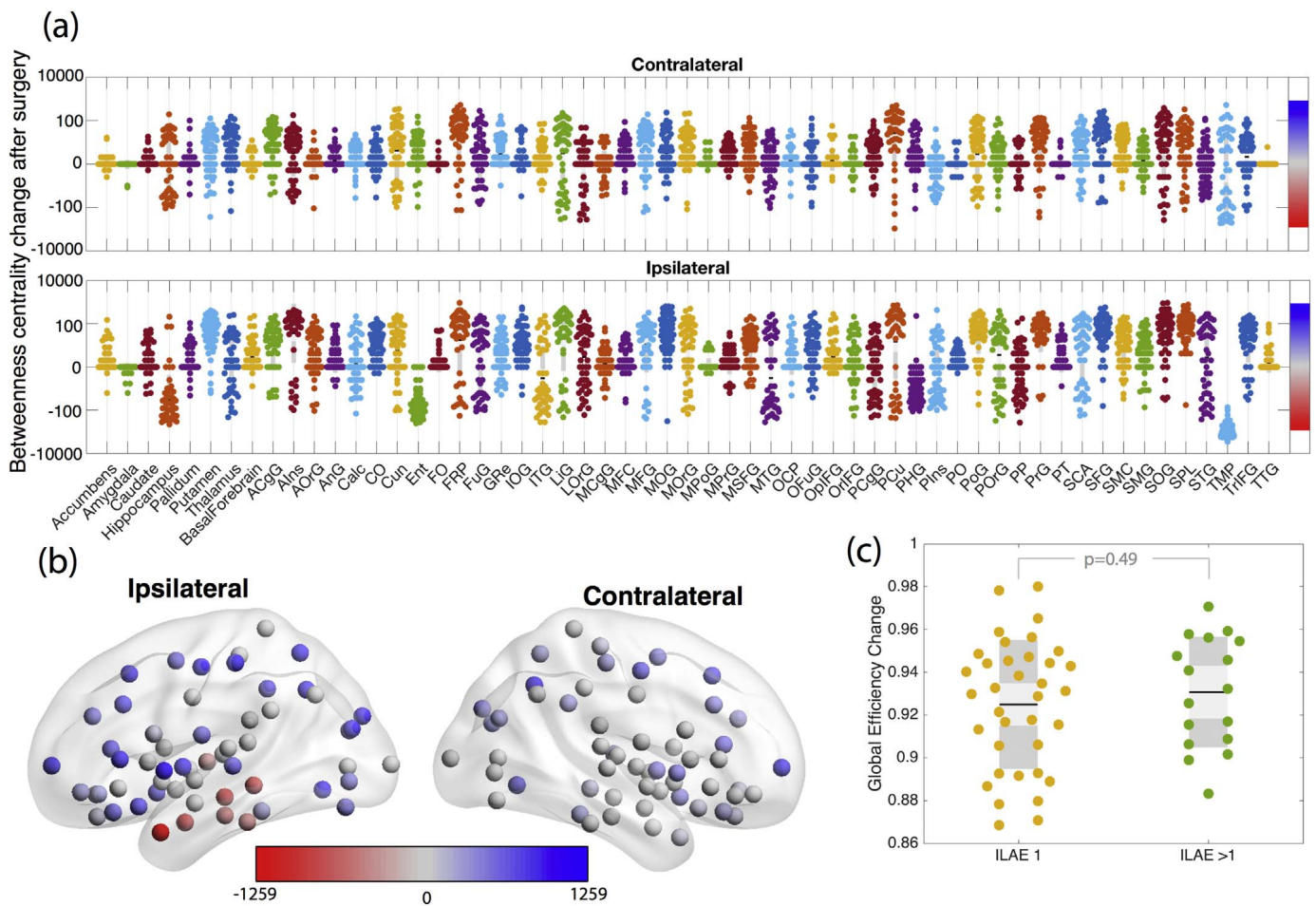

**Fig. 4.** Widespread changes in region betweenness centrality following surgery.
**(a)** Enhancements (positive values) and reductions (negative values) in betweenness centrality following surgery. A value of 0 indicates no change. **(b)** Median change of betweenness centrality shows that decreases are generally constrained to the resected regions (red) whereas increases are widespread across many other regions (blue). Colour bars in panel (b) and (a) are consistent and correspond to the change in betweenness centrality (y axis in panel a). Red (blue) indicates a decrease (increase) in betweenness centrality after surgery. **(c)** Global network efficiency is typically reduced by < 10% following surgery, and does not predict outcome.

Fig. 6 shows the median change in betweenness centrality across all subjects. Decreases in connection betweenness centrality were broadly limited to ipsilateral temporal areas whereas the increases were more widely distributed.

### 3.3. Network change association with seizure outcome

As an outlook to prospective applications, we used a machine learning strategy to retrospectively investigate the association between network changes – region volume, region strength, and connection strength – with seizure outcome.

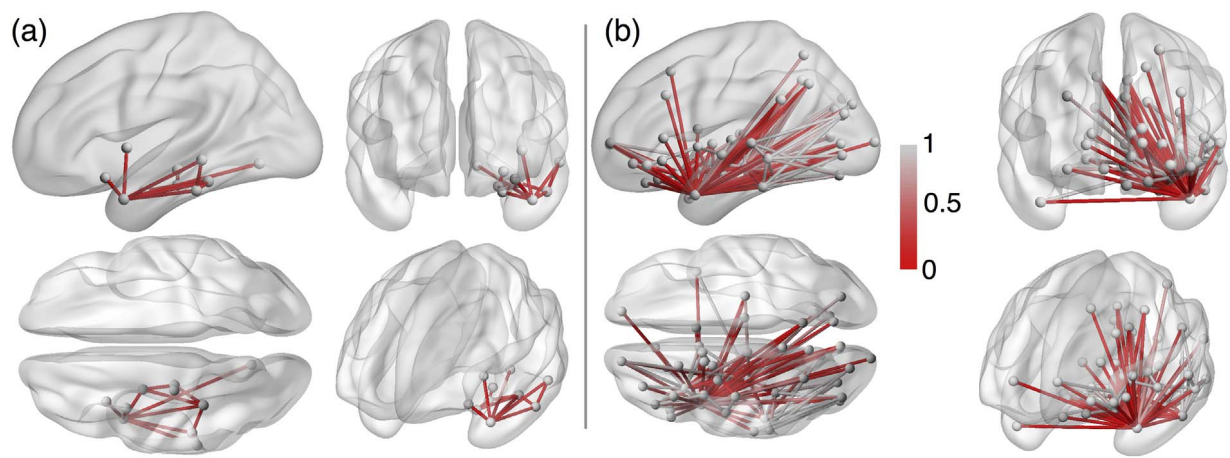

**Fig. 5.** Connectivity is disrupted by surgery.
**(a)** Connections which are reduced in strength following surgery in all patients. **(b)** Median reduction in connectivity strength across all patients. A value of 1 (0) indicates that the median change to a connection following surgery is a 0% (100%) reduction in the number of streamlines. Only values < 1 are shown and therefore shows only connections which are reduced in strength in the majority of patients (as opposed to connections which are reduced in all patients - shown in panel a). Connections in panel (b) are a superset of those in panel (a).

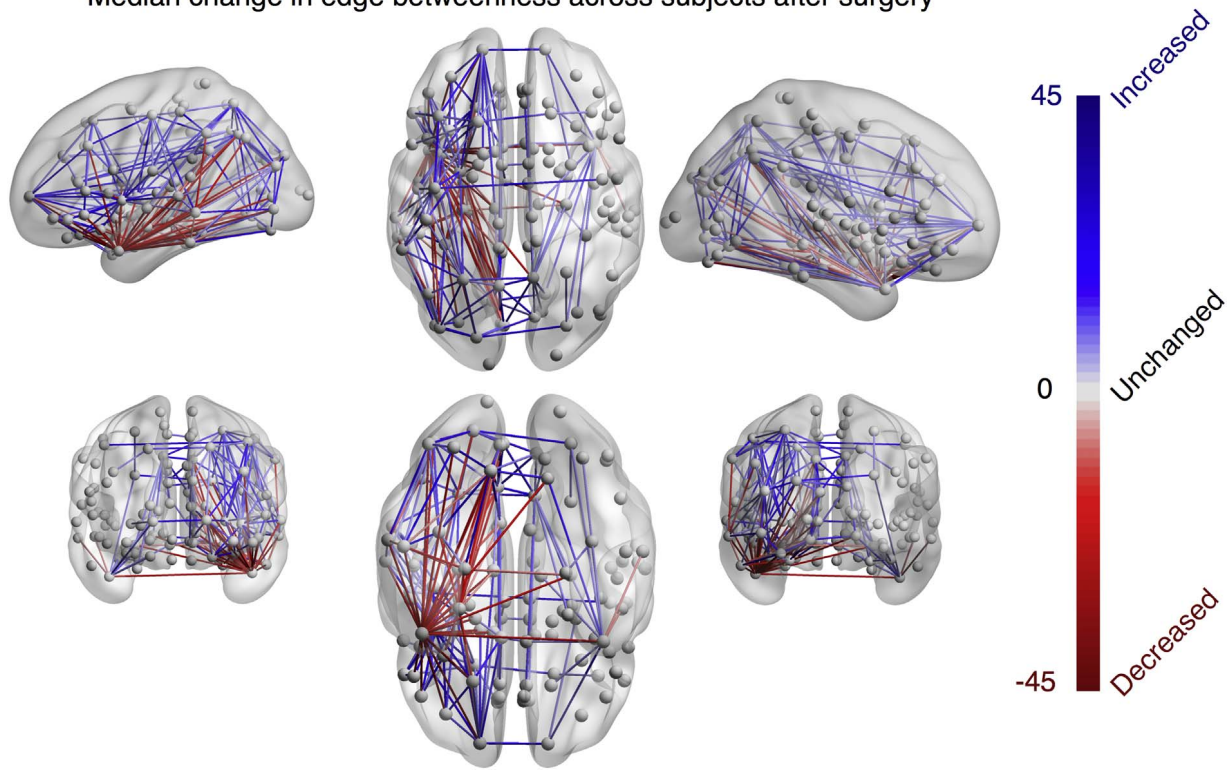

**Fig. 6.** Widespread changes in connection betweenness centrality following surgery.
Median decreases (increases) in connection betweenness centrality across all subjects shown in red (blue).

Our machine learning approach minimises the number of features that give a high accuracy of outcome classification. As input features we incorporated all changes in volume, region strength, and connection strength and applied an elastic net algorithm to select those which were most useful. Fig. 7 shows those features selected by the algorithm. Fifteen features were selected, all of which were connections. This suggests that incorporating knowledge of regional properties (strength and volume change) does not improve the prediction substantially. All fifteen connections in Fig. 7 are coloured according to the difference in mean value between groups. In other words, a negative value shows that the mean ILAE > 1 patient has a greater reduction in connectivity than the mean ILAE 1 patient (and vice-versa). The greatest reduction any group can have is 1 (i.e. a 100% reduction). Values are therefore bound between ± 1, and consequently not normally distributed. We therefore show the difference in mean, rather than the effect size. Eight of the features are connections involving the temporal pole. The list of selected features is given in Table S13 and are visualised in movie S14. Replication of these results using normalised data, preoperative data, or the alternative FreeSurfer parcellation did *not* lead to substantial improvements in accuracy (Supplementary results S10). Those who became seizure free had greater reductions of connectivity into the ipsilateral superior temporal lobe and frontal lobe.

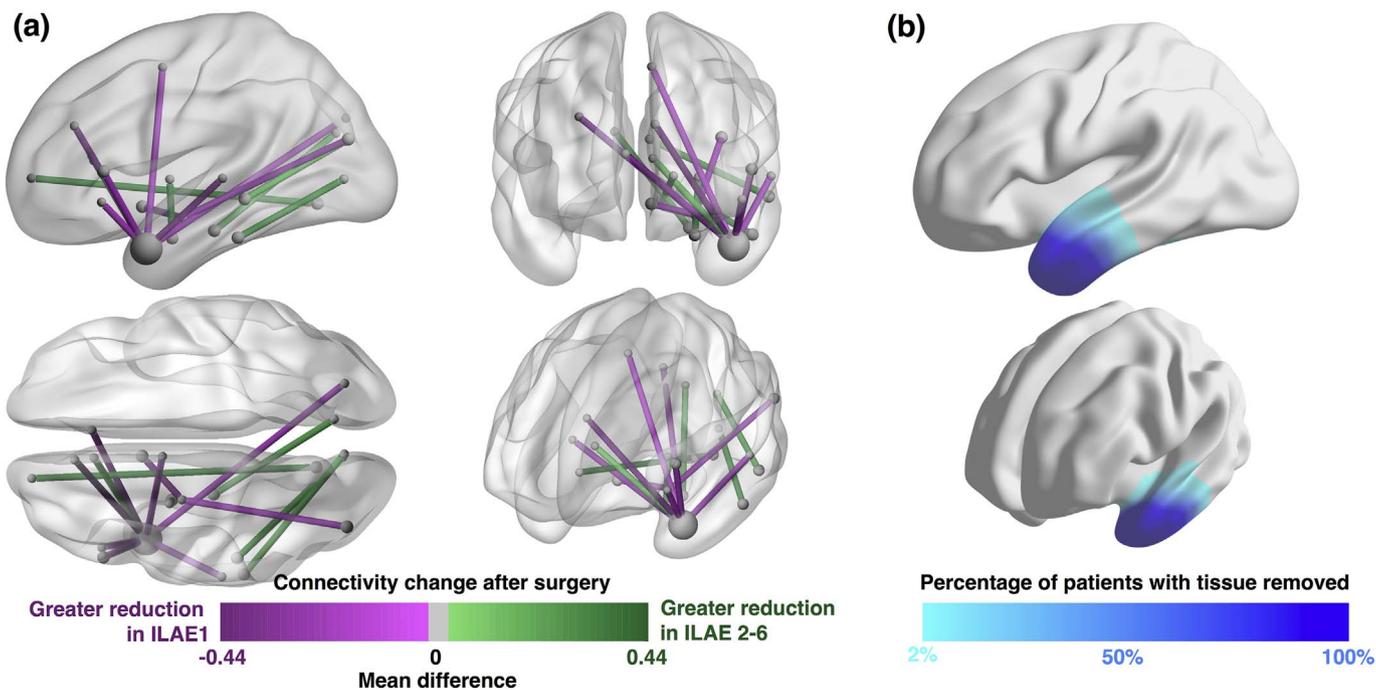

**Fig. 7.** Features derived with the elastic net algorithm.
**(a)** Fifteen features which were found to be most informative using our machine learning pipeline. All are connections which are reduced in strength following surgery. Colour coding shows the difference in mean reduction between groups. Connections in red are reduced by surgery more in seizure-free patients than in not-seizure-free patients. **(b)** Correspondence with resected tissue.

Table 3 shows our findings including accuracy, sensitivity and specificity when using these features. Our specificity of 64.7% reflects the finding that of all patients that the model predicted would have a not-seizure-free outcome, 64.7% actually had a not-seizure-free outcome. This is substantially higher than the empirical distribution (32%).

## 4. Discussion

In this study we have used detailed neuroanatomical information from both pre- and postoperative MRI to infer the impact of surgery on patient-specific brain networks. Our three main contributions are as follows.

First, we find that the impact of surgery leads to a reduction of < 10% in efficiency in the majority of patients, due to a redirection of shortest paths. Second, there is no single feature which fully accounts for outcome with 100% accuracy. This means that there is likely not a single 'target' to aim for with surgery in this patient group that is captured by our data. Third, we have demonstrated that machine learning, in conjunction with connectivity change metrics, can produce predictions with high accuracy, and specificity which is twice as high as the empirical distribution. This means that around two thirds of the patients we predict are not seizure free, are actually not seizure free. This could help preoperatively in identifying which patients might have lower chances of becoming seizure free and put extra consideration in offering surgery or when consenting those patients.

Most previous network-based studies in epilepsy typically perform group comparisons of patients relative to controls using either structural networks (Kamiya et al., 2016; Taylor et al., 2015; Lemkaddem et al., 2014; de Salvo et al., 2014; Bonilha et al., 2012), functional networks (Chavez et al., 2010; Liao et al., 2010), or both (Zhang et al., 2011; Douw et al., 2015; Wirsich et al., 2016). Few have investigated alterations in global network properties with respect to surgical outcome (although see He et al., 2017 and Morgan et al., 2017 for recent examples). Bonilha and colleagues utilised preoperative structural networks inferred from diffusion MRI to investigate network properties with respect to surgical outcomes (Bonilha et al., 2013). This includes a multi-centre study using machine learning which is perhaps most similar to our work (Munsell et al., 2015). That study analysed preoperative structural networks and compared several machine learning algorithms for feature selection and classification for prediction of surgical outcomes. They found that the best performing algorithm utilised an elastic net/support vector machine combination, and achieved a cross-site accuracy of 66%, with a same-site accuracy of 70%. We therefore chose to use broadly the same technique for classification here. Instead of investigating preoperative networks, however, we investigate the *difference* networks - i.e. the inferred changes brought about by patient-specific surgery. These difference networks have led us to achieve a same-site accuracy of 79% in our data.

We have performed a retrospective analysis of connectivity change using a mask drawn from postoperative data – i.e. the tissue that was resected. However, we envisage that in future a prospective prediction could be performed. This should be possible using the preoperative data and a mask of what the surgeon *intends* to resect (e.g. through the manual adaptation of an 'average mask' in the case of a standardised surgery such as anterior temporal lobe resection). Indeed, in theory a prediction is possible for any given presurgical resection plan. Fig. 8 suggests how this could be incorporated into such an evaluation, with the aim to address the question of, 'will this proposed resection render the patient seizure free?' We do not address the separate question of, 'what is the optimal resection for this patient?', which has to integrate the need to avoid damage to critical structures (e.g. Winston et al., 2012). This study has developed a pipeline for answering the first of those two questions, and retrospectively validated it with 79% accuracy. Approaches to investigate the second of those questions have been proposed recently (Sinha et al., 2017; Goodfellow et al., 2016).

### 4.1. Regional change following surgery

A potentially surprising result (Fig. 2), is that we found that the amount of resected tissue did not appear to predict outcome. This is in contrast to the work of (Wyler et al., 1995) and (Shamim et al., 2009) who have shown that larger resections *do* correlate with seizure outcomes. However, other studies have found conflicting results which accord with our own findings (Hardy et al., 2003; Keller et al., 2015; Liao et al., 2016). The variations in the literature may be due to different populations of left/right TLE, evidence of hippocampal sclerosis, and follow-up duration. Here we have controlled for these aspects where possible with no significant differences between outcome groups (Table 1).

Our finding that there is variation between subjects, in terms of proportion of regions removed, can be explained by two potential factors. First, there are differences in tissue removed between subjects, and second, there are differences in regions between subjects. The latter was one of our motivations for using two different parcellations to aid comparison (the second of which is shown in supplementary materials).

**Table 3**
Confusion matrix indicating performance of machine learning framework for predicting surgical outcomes.

|  |  | Actual surgical outcome | |
| --- | --- | --- | --- |
|  |  | Seizure free = 36 | Not seizure free = 17 |
| Predicted outcome | Seizure free = 37 | True positive = 31 | False positive = 6 (Type I Error) |
|  | Not seizure free = 16 | False negative = 5 (Type II Error) | True negative = 11 |
|  | Accuracy = 0.792 | True Positive Rate, or Sensitivity = 0.861 False Negative Rate, or Miss rate = 0.139 | False Positive Rate or Fall out = 0.353 True Negative Rate, or Specificity = 0.647 |

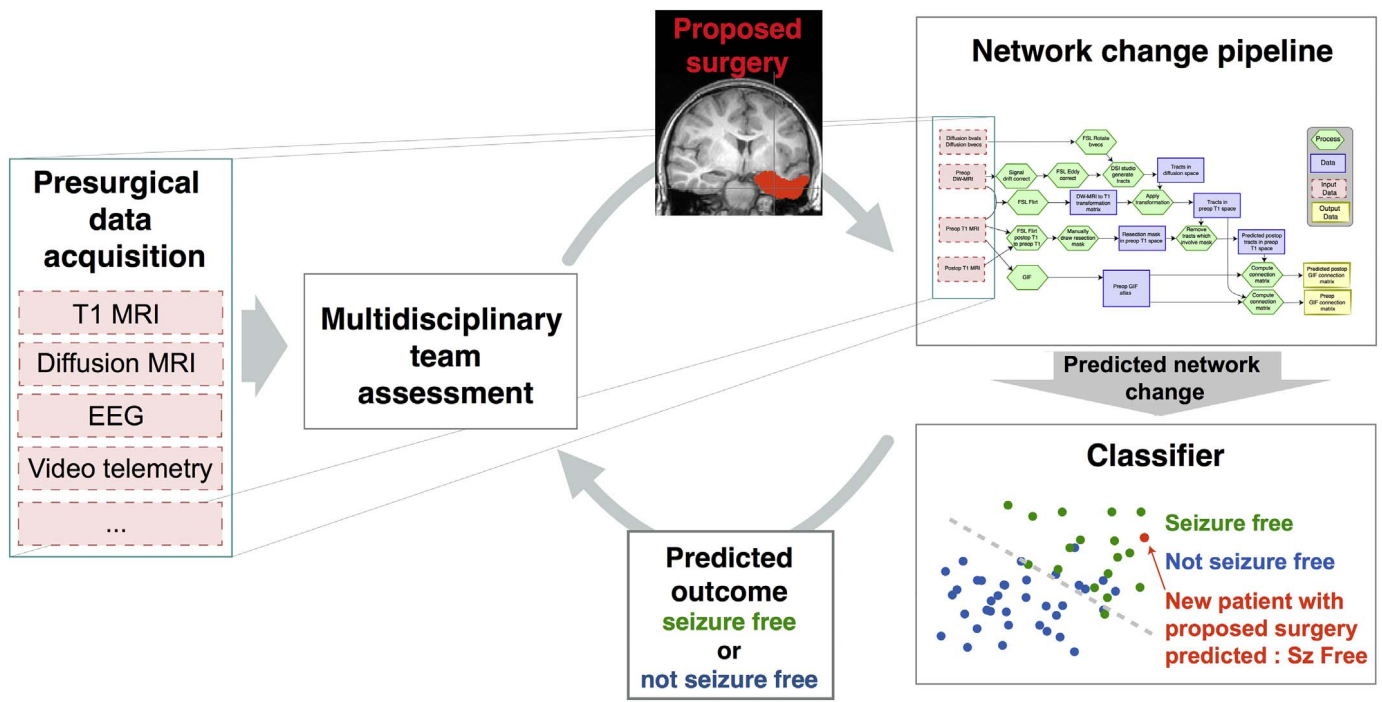

**Fig. 8.** Potential incorporation of pipeline for prospective evaluation.
Suggestion of how our approach could be used as part of the pre-surgical evaluation. From left to right: Multimodal data is acquired and evaluated at the multi-disciplinary team meeting. If surgery is recommended a proposed surgical resection mask is drawn, then fed into the connection change pipeline. The pipeline then gives a predicted post-operative network given the patient-specific mask and patient-specific presurgical DTI & T1w MRI. The pipeline uses these predicted connectivity changes, along with the pre-existing feature set to predict seizure outcome. In this example the predicted outcome would be seizure free.

Furthermore, the resolution of the parcellation (i.e. the number of regions that the brain network is divided into) will also affect results. Here we use fairly low resolution parcellations of ≈ 100 regions in contrast to recently proposed higher resolutions (Besson et al., 2014a, 2014b; Taylor et al., 2017; Besson et al., 2017). Our reasoning for this is that in this study we compare regional properties between subjects; thus it is important that we are confident that one region in one subject is the same region in another, and can be reliably reproduced. One of the reasons that the FreeSurfer software is so popular for this is that the authors have demonstrated confidence maps at region boundaries (Fischl et al., 2004). One of the drawbacks of using such low resolution connectivity matrices however is the inability to investigate detailed within-area architecture as in (Taylor et al., 2017). Indeed, this may be an important and productive area of future research.

### 4.2. Network change following surgery

We suggest that our finding of only a < 10% reduction in network efficiency following surgery (Fig. 4c) may explain why surgery usually leads to relatively few severe side effects. The reason for only a subtle change in efficiency is the redirection of shortest paths via other connections as shown in Fig. 6. We speculate that the *decreases* in edge betweenness correspond to those connections which undergo Wallerian degeneration. Furthermore, we speculate that connections with *increased* betweenness centrality will undergo 'strengthening', potentially reflected by increases in FA on structural imaging postoperatively. Qualitative evidence to support this has been shown by (Winston et al., 2014) who found decreases in FA in tracts involving the resected tissue and increased FA in more distant ipsilateral superior tracts within fairly short timescales of three months. Further support for this hypothesis comes from a study by (Jeong et al., 2016) who report increases in the FA of connections following frontal lobe epilepsy surgery far from the surgical site. Other studies have also made similar findings (Yogarajah et al., 2010; Pustina et al., 2014; Faber et al., 2013). Taken together, this suggests a potential rerouting of connectivity, via new shortest paths, leading to increased FA and thus avoiding major cognitive deficits following surgery.

The relationship between network efficiency, shortest path lengths, and function is not entirely clear. However, path transitivity, which is strongly related to path length has been shown to strongly correlate to functional connectivity (Goñi et al., 2014). Network efficiency itself has also been shown to relate to intelligence in functional networks in healthy controls (van den Heuvel et al., 2009). It remains to be seen how *changes* to network efficiency metrics following surgery relate to patient changes in brain function beyond seizure outcome.

In addition to widespread network changes such as those described above, the network change pipeline also allows the inference of more localised differences. As shown in Fig. 3, we found decreases in node strength in extratemporal areas, particularly the basal forebrain in the frontal lobe. This leads us to suggest that frontal lobe function may be altered following surgery. Several studies are in agreement with this. For example, Stretton et al. (2014) found improvements in frontal lobe working memory function after surgery, while Martin et al. (2000) showed significant increases in verbal fluency following surgery. However, other studies have also found conflicting results (see Stretton and Thompson, 2012 for review). We hypothesise these improvements are related to post-surgical alterations in connectivity, and the variability in the results may be down to patient variation (i.e. the large spread in Fig. 3). This effect is typically referred to as a process of functional improvement once nociferous cortex has been removed. Future studies correlating these changes in connectivity with changes in frontal lobe function will help to elucidate these relationships, potentially predicting not only seizure outcome, but also functional outcomes. Indeed, this has already been demonstrated in stroke where connectivity change metrics have been found to relate to functional outcomes (Kuceyeski et al., 2015a, 2015b).

### 4.3. Machine learning for outcome prediction

Although very few have used brain network data, several studies in recent years have begun to apply machine learning strategies for surgical outcome prediction. The benefits of the approach appear promising. For example, (Bernhardt et al., 2015), using hippocampal measurements as features, made outcome predictions with 81% accuracy. In another study of surgical outcomes in temporal lobe epilepsy (Memarian et al., 2015) demonstrated accuracy of 95% using various features derived from both imaging and demographic data. In addition, (Feis et al., 2013) achieved accuracy of outcome prediction of up to 96% using preoperative T1 measurements of white matter volumes. Other studies have also used machine learning algorithms, along with imaging data, to classify lateralisation of seizure onset (Focke et al., 2012; Bennett et al., 2017). Our prediction accuracy is in a broadly similar range to many of these and may be improved by including other features such as those used for generating predictive nomograms (Jehi et al., 2015).

Multiple approaches exist in machine learning to attempt to improve the accuracy of the outcome prediction by selecting the most predictive features from a high-dimensional feature space (Tibshirani, 1996, Hoerl and Kennard, 2004, Breiman et al., 1984, Breiman et al., 1984). Here we have adopted a fairly conservative approach in which we implement elastic net regularisation (Zou and Hastie, 2005) while incorporating regularisation parameters which identify a minimum number of features important for the prediction. Even using this conservative approach we obtain a prediction accuracy of 79.2%. It is worth noting that we impose a prior probability expectation of 50%. If we did not impose this, and instead used the empirical distribution of outcomes (68%: 32% - row 1 in Table 1), it would be easy to get accuracies of at least 68% by simply suggesting *all* patients will be seizure-free. The prediction accuracy of 79.2%, even using our highly conservative approach, is therefore encouraging.

*4.4. Strengths & weaknesses*

There are several strengths to this work. For example, here we have a fairly large cohort of 53 patients which were all scanned on the same scanner using the same scanning protocol. We also have good consistency in terms of operation type (all anterior temporal lobe resection performed by the same 2 surgeons using the same technique), and consistency in follow-up duration for outcome - 12 months. The consistency in follow-up duration is particularly important because outcomes can change over time (De Tisi et al., 2011). A further strength of our study, in comparison to some other network connectivity change studies, is that our analysis is performed in subject (native) space. This means nonlinear deformations of the scans to a common space, which may decrease the accuracy of the region mapping, are not necessary. Our investigation of *changes* in structural connectivity following surgery using a ChaCo-like approach is, to our knowledge, novel in epilepsy. Moreover, the ChaCo approach has the advantage of not requiring arbitrary thresholding of the network, or comparison to arbitrarily chosen random networks as a baseline – instead the patient's own pre-operative network is used.

A possible weakness of this study is that we have flipped the connectivity matrices in right TLE patients, simplifying our analysis into ipsi- and contralateral hemispheres. Although there is precedence for this approach (Bonilha et al., 2013; Munsell et al., 2015; Ji et al., 2015) it has been suggested that left and right TLE are not simply the mirror image of each other (Besson et al., 2014a, 2014b), especially with respect to post-surgical cognitive deficits and resection size which is typically larger in the non-dominant hemisphere. We did try to mitigate this however by ensuring the proportion of left:right patients in each outcome group are similar ($p = 0.3353$; Table 1).

Another potential bias is the resection mask, which was manually drawn for all subjects by the same rater, with a majority subset drawn by a second rater (results S9). Variations between raters exist, but are small. The correlation of network features between raters is > 0.98, giving confidence in the reproducibility of the masks drawn here. Our choice of tractography algorithm was influenced by the intention to limit false positive connections as suggested by Zalesky et al. (2016). Despite this, there are known limitations with all approaches and this study should also be considered with the known limitations of the imaging technologies in mind. For example, the inability to resolve connection direction and potential bias for tractography to favour shorter, straighter connections (Jones, 2010). Our use of number of streamlines as a connectivity metric is based on the extensive prior literature of ChaCo analysis by Kuceyeski et al. (2013, 2015a, 2015b). However, other connection strength metrics, such as density (Taylor et al., 2015), or volume (Irimia and Van Horn, 2016) may also yield informative results.

Our network change pipeline allows to infer the *immediate* network change brought about by surgery. It is worth highlighting that this does not necessarily represent the *long-term* brain network changes in the months/years that occur as a result of e.g. plasticity or degeneration. This is both a strength and a weakness of our approach and is one reason why we assess outcomes at only 12 months. The mechanisms underlying why some patients relapse to seizure recurrence only at several years after surgery (De Tisi et al., 2011) are unclear and would require longitudinal diffusion MRI data gathered over several years to investigate in this framework. The potential to use our approach prospectively (Fig. 8) is a distinct advantage however.

*4.5. Outlook and wider implications*

Although we have focussed exclusively on temporal lobe epilepsy surgery here, this method could be generalized to other types of neurosurgery such as those for other epilepsies such as neocortical epilepsy, tumours or movement disorders. Additionally, other outcome criteria could also be predicted. For example, instead of predicting seizure outcome, functional deficits arising due to damage to known functional subnetworks (e.g. somatomotor, attention, visual) could be predicted using existing atlases (Yeo et al., 2011). Our approach would be possible if preoperative imaging data is available and a proposed resection mask drawn before the surgery. Deficits for the given mask (surgery) could then be predicted for individual subjects. Additionally, multiple proposed masks (surgeries) could be tested against each other *in-silico*, the results for each compared, and the optimal surgery predicted. Algorithms for a similar approach are already being developed for invasive electrode recordings (Rodionov et al., 2013; Sinha et al., 2017).

*4.6. Conclusion*

It is to be expected that outcome from surgery will depend not only on the network connectivity of the brain before surgery, but the disruption to this network caused by the surgery itself. This can be measured as the *change* in connectivity, which can be predicted for a given resection preoperatively. Our analysis shows that such connectivity change metrics can be used to retrospectively predict seizure outcome with high accuracy, and may relate to functional outcomes.

Supplementary data to this article can be found online at https://doi.org/10.1016/j.nicl.2018.01.028.

**Acknowledgments**

We are grateful to the Wolfson Foundation and the Epilepsy Society for supporting the Epilepsy Society MRI scanner.

PNT was funded by Wellcome Trust (105617/Z/14/Z). PNT thanks Rob Forsyth and Helen Marley for discussion. GPW was supported by the MRC (G0802012, MR/M00841X/1). SBV is funded by the National Institute for Health Research University College London Hospitals Biomedical Research Centre (NIHR BRC UCLH/UCL High Impact Initiative).

This study was carried out in part at UCLH and the support of the NIHR-funded UCLH/UCL BRC is gratefully acknowledged. JD is supported by the NIHR Senior investigator scheme.

## References

Andersson, Jesper L.R., Sotiropoulos, Stamatios N., 2016. An integrated approach to correction for off-resonance effects and subject movement in diffusion mr imaging. NeuroImage 125, 1063–1078 (Elsevier).

Bauer, Roman, Kaiser, Marcus, 2017. Nonlinear growth: an origin of hub organization in complex networks. Royal Soc. Open Sc. 4 (3), 160691 (The Royal Society).

Bennett, O., Cardoso, J., Duncan, J.S., Winston, G., Ourselin, S., 2017. Learning how to See the Invisible - Using Machine Learning to Find Underlying Abnormality Patterns in Reportedly Normal MR Brain Images from Patients with Epilepsy. Proceedings. ISMRM.

Bernhardt, Boris C., Chen, Zhang, He, Yong, Evans, Alan C., Bernasconi, Neda, 2011. Graph-theoretical analysis reveals disrupted small-world organization of cortical thickness correlation networks in temporal lobe epilepsy. Cereb. Cortex 21 (9), 2147–2157 (Oxford Univ Press).

Bernhardt, Boris C., Hong, Seok-Jun, Bernasconi, Andrea, Bernasconi, Neda, 2015. Magnetic resonance imaging pattern learning in temporal lobe epilepsy: classification and prognostics. Ann. Neurol. 77 (3), 436–446 (Wiley Online Library).

Besson, Pierre, Dinkelacker, Vera, Valabregue, Romain, Thivard, Lionel, Leclerc, Xavier, Baulac, Michel, Sammler, Daniela, et al., 2014a. Structural connectivity differences in left and right temporal lobe epilepsy. NeuroImage 100, 135–144 (Elsevier).

Besson, Pierre, Lopes, Renaud, Leclerc, Xavier, Derambure, Philippe, Tyvaert, Louise, 2014b. Intra-subject reliability of the high-resolution whole-brain structural connectome. NeuroImage 102, 283–293 (Elsevier).

Besson, Pierre, Kathleen Bandt, S., Proix, Timothée, Lagarde, Stanislas, Jirsa, Viktor K., Ranjeva, Jean-Philippe, Bartolomei, Fabrice, Guye, Maxime, 2017. Anatomic consistencies across epilepsies: a stereotactic-EEG informed high-resolution structural connectivity study. Brain 140 (10), 2639–2652.

Betzel, Richard F., Avena-Koenigsberger, Andrea, Goñi, Joaquín, He, Ye, De Reus, Marcel A., Griffa, Alessandra, Vértes, Petra E., et al., 2016. Generative models of the human connectome. NeuroImage 124, 1054–1064 (Elsevier).

Bonilha, Leonardo, Nesland, Travis, Martz, Gabriel U., Joseph, Jane E., Spampinato, Maria V., Edwards, Jonathan C., Tabesh, Ali, 2012. Medial temporal lobe epilepsy is associated with neuronal fibre loss and paradoxical increase in structural connectivity of limbic structures. J. Neurol. Neurosurg. Psychiatry 83 (9), 903–909 (BMJ Publishing Group Ltd).

Bonilha, Leonardo, Helpern, Joseph A., Sainju, Rup, Nesland, Travis, Edwards, Jonathan C., Glazier, Steven S., Tabesh, Ali, 2013. Presurgical connectome and postsurgical seizure control in temporal lobe epilepsy. Neurology 81 (19), 1704–1710 (AAN Enterprises).

Bonilha, Leonardo, Jensen, Jens H., Baker, Nathaniel, Breedlove, Jesse, Nesland, Travis, Lin, Jack J., Drane, Daniel L., Saindane, Amit M., Binder, Jeffrey R., Kuzniecky, Ruben I., 2015. The brain connectome as a personalized biomarker of seizure outcomes after temporal lobectomy. Neurology 84 (18), 1846–1853 (AAN Enterprises).

Breiman, L., Friedman, J.H., Alshen, R.A., Stone, C.J., 1984. CART: Classification and Regression Trees. Wadsworth, Belmont, CA.

Cardoso, M. Jorge, Modat, Marc, Wolz, Robin, Melbourne, Andrew, Cash, David, Rueckert, Daniel, Ourselin, Sebastien, 2015. Geodesic information flows: spatially-variant graphs and their application to segmentation and fusion. IEEE Trans. Med. Imaging 34 (9), 1976–1988 (IEEE).

Casanova, Ramon, Wagner, Benjamin, Whitlow, Christopher, Williamson, Jeff, Shumaker, Sally, Maldjian, Joseph, Espeland, Mark, 2011. High dimensional classification of structural Mri Alzheimer's disease data based on large scale regularization. Front. Neuroinform. 5, 22. http://dx.doi.org/10.3389/fninf.2011.00022.

Chavez, Mario, Valencia, Miguel, Navarro, Vincent, Latora, Vito, Martinerie, Jacques, 2010. Functional modularity of background activities in normal and epileptic brain networks. Phys. Rev. Lett. 104 (11), 118701 (APS).

Cook, Philip A., Symms, Mark, Boulby, Philip A., Alexander, Daniel C., 2007. Optimal acquisition orders of diffusion-weighted mri measurements. J. Magn. Reson. Imaging 25 (5), 1051–1058 (Wiley Online Library).

Crossley, Nicolas A., Mechelli, Andrea, Scott, Jessica, Carletti, Francesco, Fox, Peter T., McGuire, Philip, Bullmore, Edward T., 2014. The hubs of the human connectome are generally implicated in the anatomy of brain disorders. Brain 137 (8), 2382–2395 (Oxford Univ Press).

De Tisi, Jane, Bell, Gail S., Peacock, Janet L., McEvoy, Andrew W., Harkness, William F.J., Sander, Josemir W., Duncan, John S., 2011. The long-term outcome of adult epilepsy surgery, patterns of seizure remission, and relapse: a cohort study. Lancet 378 (9800), 1388–1395 (Elsevier).

Desikan, Rahul S., Ségonne, Florent, Fischl, Bruce, Quinn, Brian T., Dickerson, Bradford C., Blacker, Deborah, Buckner, Randy L., et al., 2006. An automated labeling system for subdividing the human cerebral cortex on Mri scans into gyral based regions of interest. NeuroImage 31 (3), 968–980 (Elsevier).

Douw, Linda, DeSalvo, Matthew N., Tanaka, Naoaki, Cole, Andrew J., Liu, Hesheng, Reinsberger, Claus, Stufflebeam, Steven M., 2015. Dissociated multimodal hubs and seizures in temporal lobe epilepsy. Ann. Clin. Transl. Neurol. 2 (4), 338–352 (Wiley Online Library).

Duchi, John, Shalev-Shwartz, Shai, Singer, Yoram, Chandra, Tushar, 2008. Efficient projections onto the L1-ball for learning in high dimensions. In: Proceedings of the 25th International Conference on Machine Learning. ACM, New York, NY, USA, pp. 272–279. (ICML '08). https://doi.org/10.1145/1390156.1390191.

Estrada, E., Hatano, N., 2008. Communicability in complex networks. Phys. Rev. E 77 (3), 036111.

Faber, Jennifer, Schoene-Bake, Jan-Christoph, Trautner, Peter, Lehe, Marec, Elger, Christian E., Weber, Bernd, 2013. Progressive fiber tract affections after temporal lobe surgery. Epilepsia 54 (4), e53-e57.

Feis, Delia-Lisa, Schoene-Bake, Jan-Christoph, Elger, Christian, Wagner, Jan, Tittgemeyer, Marc, Weber, Bernd, 2013. Prediction of post-surgical seizure outcome in left mesial temporal lobe epilepsy. Neurol. Clin. 2, 903–911 (Elsevier).

Fischl, Bruce, 2012. FreeSurfer. NeuroImage 62 (2), 774–781 (Elsevier).

Fischl, Bruce, Salat, David H., Busa, Evelina, Albert, Marilyn, Dieterich, Megan, Haselgrove, Christian, van der Kouwe, Andre, et al., 2002. Whole brain segmentation: automated labeling of neuroanatomical structures in the human brain. Neuron 33 (3), 341–355 (Elsevier).

Fischl, Bruce, van der Kouwe, André, Destrieux, Christophe, Halgren, Eric, Ségonne, Florent, Salat, David H., Busa, Evelina, et al., 2004. Automatically parcellating the human cerebral cortex. Cereb. Cortex 14 (1), 11–22 (Oxford Univ Press).

Focke, Niels K., Yogarajah, Mahinda, Symms, Mark R., Gruber, Oliver, Paulus, Walter, Duncan, John S., 2012. Automated Mr image classification in temporal lobe epilepsy. NeuroImage 59 (1), 356–362 (Elsevier).

Goñi, J., van den Heuvel, M.P., Avena-Koenigsberger, A., Velez de Mendizabal, N., Betzel, R.F., Griffa, A., ... Sporns, O., 2014. Resting-brain functional connectivity predicted by analytic measures of network communication. Proc. Natl. Acad. Sci. U. S. A. 111 (2), 833–838. http://dx.doi.org/10.1073/pnas.1315529111.

Goodfellow, M., Rummel, C., Abela, E., Richardson, M.P., Schindler, K., Terry, J.R., 2016. Estimation of brain network ictogenicity predicts outcome from epilepsy surgery. Sci. Rep. 6.

Gooneratne, I.K., Mannan, S., de Tisi, J., Gonzalez, J.C., McEvoy, A.W., Miserocchi, A., Diehl, B., Wehner, T., Bell, G.S., Sander, J.W., Duncan, J.S., 2017. Somatic complications of epilepsy surgery over 25 years at a single center. Epilepsy Res. 132, 70–77.

Hardy, Steven G., Miller, John W., Holmes, Mark D., Born, Donald E., Ojemann, George A., Dodrill, Carl B., Hallam, Danial K., 2003. Factors predicting outcome of surgery for intractable epilepsy with pathologically verified mesial temporal sclerosis. Epilepsia 44 (4), 565–568 (Wiley Online Library).

He, X., Doucet, G.E., Pustina, D., Sperling, M.R., Sharan, A.D., Tracy, J.I., 2017. Presurgical thalamic "hubness" predicts surgical outcome in temporal lobe epilepsy. Neurology 10–1212.

van den Heuvel, M.P., Stam, C.J., Kahn, R.S., Pol, H.E.H., 2009. Efficiency of functional brain networks and intellectual performance. J. Neurosci. 29 (23), 7619–7624.

Hoerl, A.E., Kennard, R.W., 2004. Ridge Regression. 9780471667193 John Wiley and Sons, Inc.http://dx.doi.org/10.1002/0471667196.ess2280.

Irimia, Andrei, Van Horn, John Darrell, 2016. Scale-dependent variability and quantitative regimes in graph-theoretic representations of human cortical networks. Brain Connect. 6 (2), 152–163.

Jehi, L., Yardi, R., Chagin, K., Tassi, L., Russo, G.L., Worrell, G., ... Chauvel, P., 2015. Development and validation of nomograms to provide individualised predictions of seizure outcomes after epilepsy surgery: a retrospective analysis. Lancet Neurol. 14 (3), 283–290.

Jenkinson, Mark, Smith, Stephen, 2001. A global optimisation method for robust affine registration of brain images. Med. Image Anal. 5 (2), 143–156 (Elsevier).

Jenkinson, Mark, Bannister, Peter, Brady, Michael, Smith, Stephen, 2002. Improved optimization for the robust and accurate linear registration and motion correction of brain images. NeuroImage 17 (2), 825–841 (Elsevier).

Jenkinson, Mark, Beckmann, Christian F., Behrens, Timothy E.J., Woolrich, Mark W., Smith, Stephen M., 2012. Fsl. NeuroImage 62 (2), 782–790 (Elsevier).

Jeong, Jeong-Won, Asano, Eishi, Juhász, Csaba, Behen, Michael E., Chugani, Harry T., 2016. Postoperative axonal changes in the contralateral hemisphere in children with medically refractory epilepsy: a longitudinal diffusion tensor imaging connectome analysis. Hum. Brain Mapp. 37 (11), 3946–3956 (Wiley Online Library).

Ji, Gong-Jun, Zhang, Zhiqiang, Xu, Qiang, Wei, Wang, Jue, Wang, Zhengge, Yang, Fang, et al., 2015. Connectome reorganization associated with surgical outcome in temporal lobe epilepsy. Medicine 94 (40), e1737 (LWW).

Jones, D.K., 2010. Challenges and limitations of quantifying brain connectivity in vivo with diffusion MRI. Quant. Imaging Med. Surg. 2, 341–355.

Jutila, L., Immonen, A., Mervaala, E., Partanen, J., Partanen, K., Puranen, M., Kälviäinen, R., et al., 2002. Long term outcome of temporal lobe epilepsy surgery: analyses of 140 consecutive patients. J. Neurol. Neurosurg. Psychiatry 73 (5), 486–494 (BMJ Publishing Group Ltd).

Kamiya, Kouhei, Amemiya, Shiori, Suzuki, Yuichi, Kunii, Naoto, Kawai, Kensuke, Akira, Kunimatsu, Nobuhito, Saito, Shigeki, Aoki, Kuni, Ohtomo, 2016. Machine learning of Dti structural brain connectomes for lateralization of temporal lobe epilepsy. Magn. Reson. Med. Sci. 15 (1), 121–129 (Japanese Society for Magnetic Resonance in Medicine).

Keller, Simon S., Richardson, Mark P., Schoene-Bake, Jan-Christoph, O'muircheartaigh, Jonathan, Elkommos, Samia, Kreilkamp, Barbara, Goh, Yee Yen, Marson, Anthony G., Elger, Christian, Weber, Bernd, 2015. Thalamotemporal alteration and postoperative seizures in temporal lobe epilepsy. Ann. Neurol. 77 (5), 760–774 (Wiley Online Library).

Kuceyeski, Amy, Maruta, Jun, Relkin, Norman, Raj, Ashish, 2013. The network modification (nemo) tool: elucidating the effect of white matter integrity changes on cortical and subcortical structural connectivity. Brain Connect. 3 (5), 451–463 (Mary Ann Liebert, Inc. 140 Huguenot Street, 3rd Floor New Rochelle, NY 10801 USA).

Kuceyeski, A.F., Vargas, W., Dayan, M., Monohan, E., Blackwell, C., Raj, A., Fujimoto, K., Gauthier, S.A., 2015a. Modeling the relationship among gray matter atrophy, abnormalities in connecting white matter, and cognitive performance in early multiple sclerosis. Am. J. Neuroradiol. 36 (4), 702–709 (Am Soc Neuroradiology).

Kuceyeski, Amy, Navi, Babak B., Kamel, Hooman, Relkin, Norman, Villanueva, Mark, Raj,

Ashish, Toglia, Joan, O'dell, Michael, Iadecola, Costantino, 2015b. Exploring the brain's structural connectome: a quantitative stroke lesion-dysfunction mapping study. Hum. Brain Mapp. 36 (6), 2147–2160 (Wiley Online Library).

Kuceyeski, Amy, Navi, Babak B., Kamel, Hooman, Raj, Ashish, Relkin, Norman, Toglia, Joan, Iadecola, Costantino, O'dell, Michael, 2016. Structural connectome disruption at baseline predicts 6-months post-stroke outcome. Hum. Brain Mapp. 37 (7), 2587–2601.

Leemans, Alexander, Jones, Derek K., 2009. The B-matrix must be rotated when correcting for subject motion in DTI data. Magn. Reson. Med. 61 (6), 1336–1349.

Lemkaddem, Alia, Daducci, Alessandro, Kunz, Nicolas, Lazeyras, François, Seeck, Margitta, Thiran, Jean-Philippe, Vulliémoz, Serge, 2014. Connectivity and tissue microstructural alterations in right and left temporal lobe epilepsy revealed by diffusion spectrum imaging. Neuroimaging Clin. N. Am. 5, 349–358 (Elsevier).

Liao, Wei, Zhang, Zhiqiang, Pan, Zhengyong, Mantini, Dante, Ding, Jurong, Duan, Xujun, Cheng, Luo, Lu, Guangming, Chen, Huafu, 2010. Altered functional connectivity and small-world in mesial temporal lobe epilepsy. PLoS One 5 (1), e8525 (Public Library of Science).

Liao, Wei, Ji, Gong-Jun, Xu, Qiang, Wei, Wei, Wang, Jue, Wang, Zhengge, Yang, Fang, et al., 2016. Functional connectome before and following temporal lobectomy in mesial temporal lobe epilepsy. Sci. Rep. 6, 23153 (Nature Publishing Group).

Liu, Jun, Ye, Jieping, 2009. Efficient Euclidean projections in linear time. In: Proceedings of the 26th Annual International Conference on Machine Learning. ACM, New York, NY, USA, pp. 657–664. ICML '09. https://doi.org/10.1145/1553374.1553459.

Liu, J., Ji, S., Ye, J., 2009a. SLEP: Sparse Learning with Efficient Projections. Arizona State University. http://www.public.asu.edu/~jye02/Software/SLEP.

Liu, Jun, Chen, Jianhui, Ye, Jieping, 2009b. Large-scale sparse logistic regression. In: Proceedings of the 15th Acm Sigkdd International Conference on Knowledge Discovery and Data Mining. ACM, New York, NY, USA, pp. 547–556. KDD '09. https://doi.org/10.1145/1557019.1557082.

Liu, Min, Chen, Zhang, Beaulieu, Christian, Gross, Donald W., 2014. Disrupted anatomic white matter network in left mesial temporal lobe epilepsy. Epilepsia 55 (5), 674–682 (Wiley Online Library).

Martin, R.C., Sawrie, S.M., Edwards, R., Roth, D.L., Faught, E., Kuzniecky, R.I., Morawetz, R.B., Gilliam, F.G., 2000. Investigation of executive function change following anterior temporal lobectomy: selective normalization of verbal fluency. Neuropsychology 14 (4), 501.

Maslov, Sergei, Sneppen, Kim, 2002. Specificity and stability in topology of protein networks. Science 296 (5569), 910–913 (American Association for the Advancement of Science).

Memarian, Negar, Kim, Sally, Dewar, Sandra, Engel, Jerome, Staba, Richard J., 2015. Multimodal data and machine learning for surgery outcome prediction in complicated cases of mesial temporal lobe epilepsy. Comput. Biol. Med. 64, 67–78 (Elsevier).

Morgan, Victoria L., Englot, Dario J., Rogers, Baxter P., Landman, Bennett A., Cakir, Ahmet, Abou-Khalil, Bassel W., Anderson, Adam W., 2017. Magnetic resonance imaging connectivity for the prediction of seizure outcome in temporal lobe epilepsy. Epilepsia 58 (7), 1251–1260.

Munsell, Brent C., Wee, Chong-Yaw, Keller, Simon S., Weber, Bernd, Elger, Christian, da Silva, Laura Angelica Tomaz, Nesland, Travis, Styner, Martin, Shen, Dinggang, Bonilha, Leonardo, 2015. Evaluation of machine learning algorithms for treatment outcome prediction in patients with epilepsy based on structural connectome data. NeuroImage 118, 219–230. http://dx.doi.org/10.1016/j.neuroimage.2015.06.008.

Osuna, Edgar, Freund, Robert, Girosi, Federico, 1997. Support vector machines: training and applications. Libr. Technol. Rep. AIM-1602.

Pustina, Dorian, Doucet, Gaelle, Evans, James, Sharan, Ashwini, Sperling, Michael, Skidmore, Christopher, Tracy, Joseph, 2014. Distinct types of white matter changes are observed after anterior temporal lobectomy in epilepsy. PLoS One 9 (8), e104211.

Rodionov, Roman, Vollmar, Christian, Nowell, Mark, Miserocchi, Anna, Wehner, Tim, Micallef, Caroline, Zombori, Gergely, et al., 2013. Feasibility of multimodal 3d neuroimaging to guide implantation of intracranial eeg electrodes. Epilepsy Res. 107 (1), 91–100 (Elsevier).

Rubinov, Mikail, Sporns, Olaf, 2010. Complex network measures of brain connectivity: uses and interpretations. NeuroImage 52 (3), 1059–1069 (Elsevier).

de Salvo, Matthew N., Douw, Linda, Tanaka, Naoaki, Reinsberger, Claus, Stufflebeam, Steven M., 2014. Altered structural connectome in temporal lobe epilepsy. Radiology 270 (3), 842–848. http://dx.doi.org/10.1148/radiol.13131044.

Shamim, Sadat, Wiggs, Edythe, Heiss, John, Sato, Susumu, Liew, Clarissa, Solomon, Jeffrey, Theodore, William H., 2009. Temporal lobectomy: resection volume, neuropsychological effects, and seizure outcome. Epilepsy Behav. 16 (2), 311–314 (Elsevier).

Sinha, N., Dauwels, J., Kaiser, M., Cash, S.S., Brandon Westover, M., Wang, Y., Taylor, P.N., 2017. Predicting neurosurgical outcomes in focal epilepsy patients using computational modelling. Brain 140 (2), 319–332.

Stretton, J., Thompson, P.J., 2012. Frontal lobe function in temporal lobe epilepsy. Epilepsy Res. 98 (1), 1–13.

Stretton, J., Sidhu, M.K., Winston, G.P., Bartlett, P., McEvoy, A.W., Symms, M.R., Koepp, M.J., Thompson, P.J., Duncan, J.S., 2014. Working memory network plasticity after anterior temporal lobe resection: a longitudinal functional magnetic resonance imaging study. Brain 137 (5), 1439–1453.

Taylor, Peter N., Han, Cheol E., Schoene-Bake, Jan-Christoph, Weber, Bernd, Kaiser, Marcus, 2015. Structural connectivity changes in temporal lobe epilepsy: spatial features contribute more than topological measures. Neuroimaging Clin. N. Am. 8, 322–328 (Elsevier).

Taylor, Peter N., Wang, Yujiang, Kaiser, Marcus, 2017. Within brain area tractography suggests local modularity using high resolution connectomics. Sci. Rep. 7 (Nature Publishing Group).

Tibshirani, Robert, 1996. Regression shrinkage and selection via the lasso. J. R. Stat. Soc. Ser. B Methodol. 267–288.

Tournier, J., Calamante, F., Connelly, A., 2012. MRtrix: diffusion tractography in crossing fiber regions. Int. J. Imaging Syst. Technol. 22 (1), 53–66.

Veropoulos, Konstantinos, Campbell, Colin, Cristianini, Nello, 1999. Controlling the sensitivity of support vector machines. In: Proceedings of the International Joint Conference on AI, pp. 55–60.

Vos, Sjoerd B., Tax, Chantal M.W., Luijten, Peter R., Ourselin, Sebastien, Leemans, Alexander, Froeling, Martijn, 2016. The importance of correcting for signal drift in diffusion Mri. In: Magnetic Resonance in Medicine. Wiley Online Library.

Wang, Zuoguan, Gunduz, Aysegul, Brunner, Peter, Ritaccio, Anthony L., Ji, Qiang, Schalk, Gerwin, 2012. Decoding onset and direction of movements using electrocorticographic (ECoG) signals in humans. Front. Neuroeng. 5.

Wheeler-Kingshott, Claudia A.M., Hickman, Simon J., Parker, Geoffrey J.M., Ciccarelli, Olga, Symms, Mark R., Miller, David H., Barker, Gareth J., 2002. Investigating cervical spinal cord structure using axial diffusion tensor imaging. NeuroImage 16 (1), 93–102 (Elsevier).

Wieser, H.G., Blume, W.T., Fish, D., Goldensohn, E., Hufnagel, A., King, D., Sperling, M.R., Lüders, H., Pedley, Timothy A., 2001. Proposal for a new classification of outcome with respect to epileptic seizures following epilepsy surgery. Epilepsia 42 (2), 282–286 (Wiley Online Library).

Winston, G.P., Daga, P., Stretton, J., Modat, M., Symms, M.R., McEvoy, A.W., Ourselin, S., Duncan, J.S., 2012. Optic radiation tractography and vision in anterior temporal lobe resection. Ann. Neurol. 71 (3), 334–341.

Winston, Gavin P., Stretton, Jason, Sidhu, Meneka K., Symms, Mark R., Duncan, John S., 2014. Progressive white matter changes following anterior temporal lobe resection for epilepsy. Neuroimaging Clin. N. Am. 4, 190–200 (Elsevier).

Wirsich, Jonathan, Perry, Alistair, Ridley, Ben, Proix, Timothée, Golos, Mathieu, Bénar, Christian, Ranjeva, Jean-Philippe, et al., 2016. Whole-brain analytic measures of network communication reveal increased structure-function correlation in right temporal lobe epilepsy. Neuroimaging Clin. N. Am. 11, 707–718 (Elsevier).

Wu, Xiaoyun, Rohini, Srihari, 2004. Incorporating prior knowledge with weighted margin support vector machines. In: Proceedings of the Tenth ACM SIGKDD International Conference on Knowledge Discovery and Data Mining. ACM.

Wyler, Allen R., Hermann, Bruce P., Somes, Grant, 1995. Extent of medial temporal resection on outcome from anterior temporal lobectomy: a randomized prospective study. Neurosurgery 37 (5), 982–991 (LWW).

Xia, Mingrui, Wang, Jinhui, He, Yong, 2013. BrainNet viewer: a network visualization tool for human brain connectomics. PLoS One 8 (7), e68910 (Public Library of Science).

Yeh, Fang-Cheng, Wedeen, Van Jay, Tseng, W.-Y.I., 2010. Generalized-sampling imaging. Med. Imaging IEEE Trans. 29 (9), 1626–1635 (IEEE).

Yeh, Fang-Cheng, Verstynen, Timothy D., Wang, Yibao, Fernández-Miranda, Juan C., Tseng, Wen-Yih Isaac, 2013. Deterministic diffusion fiber tracking improved by quantitative anisotropy. PLoS One 8 (11), e80713 (Public Library of Science).

Yeo, B.T. Thomas, Krienen, Fenna M., Sepulcre, Jorge, Sabuncu, Mert R., Lashkari, Danial, Hollinshead, Marisa, Roffman, Joshua L., et al., 2011. The organization of the human cerebral cortex estimated by intrinsic functional connectivity. J. Neurophysiol. 106 (3), 1125–1165.

Yogarajah, M., Focke, N.K., Bonelli, S.B., Thompson, P., Vollmar, C., McEvoy, A.W., Alexander, D.C., Symms, M.R., Koepp, M.J., Duncan, J.S., 2010. The structural plasticity of white matter networks following anterior temporal lobe resection. Brain 133 (8), 2348–2364.

Zalesky, Andrew, Fornito, Alex, Cocchi, Luca, Gollo, Leonardo L., van den Heuvel, Martijn P., Breakspear, Michael, 2016. Connectome sensitivity or specificity: which is more important? NeuroImage 142, 407–420.

Zhang, Z., Liao, W., Chen, H., Mantini, D., Ding, J.R., Xu, Q., Wang, Z., et al., 2011. Altered functional–structural coupling of large-scale brain networks in idiopathic generalized epilepsy. Brain 134 (10), 2912–2928 (Oxford Univ Press).

Zou, Hui, Hastie, Trevor, 2005. Regularization and variable selection via the elastic net. J. R. Stat. Soc. Ser. B (Stat Methodol.) 67 (2), 301–320. (Royal Statistical Society, Wiley). http://www.jstor.org/stable/3647580.